# A Technical Perspective on Net Neutrality

*William P. Wagner IV, Claremont Graduate University*
*william.wagner@cgu.edu*

**Abstract**

*This paper serves as a brief technical examination of Net Neutrality and the Internet fundamentals relevant to the discussion. This document seeks to provide sufficient technical perspective that it may inform the political and economic debate surrounding the issue in the United States. Further, this research demonstrates that existing Internet economics are based strictly on usage, and that this model can account for all uses. Finally, I will argue that there should be some legislation and regulation of ISPs with regard to Net Neutrality in the U.S.*

## 1. Introduction

Net Neutrality has been in the news over the last decade as a topic of political, economic, and civil rights consideration in the United States [1][2][3]. As documented in the rest of this paper, The U.S. Government as well as those of other nations [4], state and local governments, the national private networks that make up the Internet backbone, and major technology and content providers are all engaged in an active push-pull to define internet standards that protect innovation, corporate interests, and consumer needs.

The Internet is fundamental to success in the modern world [5]. Communities, law enforcement, hospitals, militaries, corporations, and individuals with a certain level of access to the internet are all demonstrably better equipped for success in the current United States and globally.

In 1996, in section 230(b) of the Communications Act of 1934, as amended, Congress describes its national Internet policy. Specifically, Congress states that it is the policy of the United States "to preserve the vibrant and competitive free market that presently exists for the Internet" [6] and "to promote the continued development of the Internet [7]."

Because of this need, it has become desirable for infrastructure planning and budgeting to define minimum standards. What those minimum standards are and what mechanism should regulate them is the first piece of the Net Neutrality debate. Proponents of Net Neutrality want legislation that prevents ISPs from restricting or prioritizing traffic – known as "fast lanes" – claiming this will stifle free-market innovation, investment, expansion, and user choice along the "Edge" of the Internet. Opponents of Net Neutrality claim the legislation will stifle free-market innovation and expansion by ISPs.

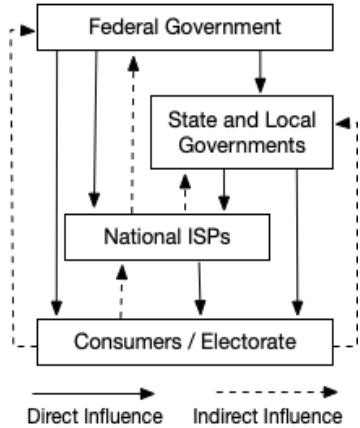

**Figure 1. Government, ISP and Consumer Relationships**

As illustrated in **fig.1**, the consumer's interests are protected through a combination of free enterprise and legislation, so this paper explores any relevant technical considerations and existing protocols, before moving to the economic models, then finally the legal fight in the courts.

This research paper demonstrates how all Internet traffic uses a process known as *Encapsulation* to achieve complete separation between the content/application developers, and the hardware/networking engineers.

Finally, this paper explains how the current economic model, based strictly on usage, accounts for **all** internet traffic down to the smallest possible unit of measure – a single bit.

It is beyond the scope of this paper to fully explore the socio-political and economic ramifications of any changes to the current system such as supply-side assistance that would encourage network expansion vs. demand-side stimulus to reduce consumer cost, however this researcher will provide some legal history and make some observations on the current state of the political and legal debate.

## 2. Definition of Key Terms

### 2.1. The Internet

It is relevant and helpful to a discussion on Net Neutrality to understand exactly what the Internet is – and note that there is little substantive difference in these definitions:

> The Internet is the global system of interconnected computer networks that use the Internet protocol suite (TCP/IP) to link devices worldwide. It is a network of networks that consists of private, public, academic, business, and government networks of local to global scope, linked by a broad array of electronic, wireless, and optical networking technologies.
> -Wikipedia

> An electronic communications network that connects computer networks and organizational computer facilities around the world —used with the except when being used attributively.
> - Merriam-Webster Dictionary

The Internet is defined as a network of networks and there is no mention of hardware, network owner, or content provider. There is no single ruling body of the Internet. See **Appendix A**.

In technical terms, we can say the Internet is by definition hardware, network, and content agnostic. These three ideas are fundamental to the internet to provide both forward and backward compatibility. The Internet does not care whether a given network uses fiber, copper or satellite. The Internet does not care whether a given user is using a Mac, PC, Smart Phone or an automated water sensor in a farmer's field. And for purposes of prioritization, the Internet protocols treat all packets of a given type equally – regardless of content, source, or destination.

This paper will use the following definition, which includes reference to the IP protocols that define the technical standards of the Internet:

**The Internet: A global system of public and private networks, a network of networks, which is hardware, network and content agnostic and based on the Internet Protocol Suite.**

### 2.2. Net Neutrality

And, of course the definition of Net Neutrality is seemingly fundamental to any discussion of Net Neutrality. Yet almost immediately we see variance in the definitions:

> Net neutrality is the principle that Internet service providers treat all data on the Internet equally, and not discriminate or charge differently by user, content, website, platform, application, type of attached equipment, or method of communication.
> - Wikipedia

> The idea, principle, or requirement that Internet service providers should or must treat all Internet data as the same regardless of its kind, source, or destination.
> - Merriam-Webster Dictionary

In this researcher's opinion, both these definitions seek to express similar ideals, but neither fully succeeds on a technical level. The Wiki definition includes a broad range of internet experiences that one might argue should not be included in a practical definition - like "user" or "method of communication." Both these aspects are superfluous to a definition of Net Neutrality. By way of illustration, the current system already tiers costs based on the type of user (ex: corporate vs. consumer or private vs government) and it tiers costs based on things like satellite vs terrestrial communication methods.

The Merriam-Webster definition hits much closer to the mark, however due to the technical nature of this discussion, I have changed the term "kind." As discussed later, there is some prioritization of data on the internet based strictly on technical specifications of TCP/IP, and the use of the term "kind" may overlap with that discussion where the terms "type" and "kind" are often used interchangeably.

Therefore, to avoid confusion, for purposes of this paper we will use the following modified definition of Net Neutrality (substituting "content" for "kind"):

**Net Neutrality: The idea, principle, or requirement that Internet service providers should or must treat all Internet data as the same regardless of its content, source, or destination.**

### 2.5. Internet Society – ISOC

The internet has grown organically, with standards that are open enough that anyone may participate, while being well defined enough that engineers can design hardware and software with confidence. Although there is no formal governing body of the Internet, the protocols that define the Internet are defined and agreed upon by an international consortium of groups, the largest of which is the Internet Society (ISOC). Within the ISOC and the other global standards groups are various open-membership advisory groups, standards groups, and definitive naming and allocation groups for things like domain names and IP addresses that require global consensus in order to operate correctly. See **Appendix A**. The group primarily responsible for establishing the technical standards of the Internet is the Internet Engineering Task Force (IETF) [9].

## 3. Fundamentals of Internet Operation

### 3.1. Hardware Independence.

As long as the hardware meets some basic technical requirements to communicate with its neighbors, there are no specific restrictions on hardware or communications medium. The Internet is Hardware Agnostic. All decisions about a given network's hardware are left to the network administrator or owner to provide the best solution for their application. Copper, fiber, satellite or cellular – that decision is up to the network owner – the Internet does not care.

### 3.2. Network Independence.

As long as the data packets arriving and leaving a network meet some basic technical requirements to communicate with their neighbors, there are no specific restrictions. The Internet does not care who the owner or administrator of a given network is or what topology that network employs – it will route packets based on "Best Effort" routing – the Internet is Network Agnostic.

### 3.3. Content Independence.

As long as the data packets leaving a network meet some basic technical requirements to communicate with their neighbors, they will be prioritized by type not content or source. For example: A home user's VoIP packet should almost always be more important than an FTP request from a big corporation, except on a subnet of only that company's FTP servers. The Internet is necessarily Content Agnostic.

## 4. Encapsulation

All of this works through a process known as **Encapsulation [Fig. 2]**. All packets sent over the internet are encapsulated. In the TCP/IP Model, there are four layers of encapsulation on the Internet.

### 4.1. Network Access Layer

The Network Access Layer (sometimes called Media Access Control or MAC) is the lowest layer, or Layer 1 in the model. This layer allows network activity at the most basic level and only for a single local network. It works at the network card level.

### 4.2. Internet Layer

Layer 2 (RFC 791, Sec. 2.2.) of the TCP/IP stack, this layer controls network activity between networks and on a global scale. It defines things like IP addresses.

### 4.3. Transport Layer

Layer 3 of the IP controls the transfer of data. After the previous layers have done their jobs of *determining the address* where a given packet should arrive, the Transport layer does the job of actually *transferring the data and confirming it was sent and received correctly*.

### 4.4. Application Layer

Layer 4 of the IP is the Application Layer. This might be a web browser or email program, Facebook or Twitter, or it might be a wind sensor or a fitness app on a watch.

### 4.5. How It Works

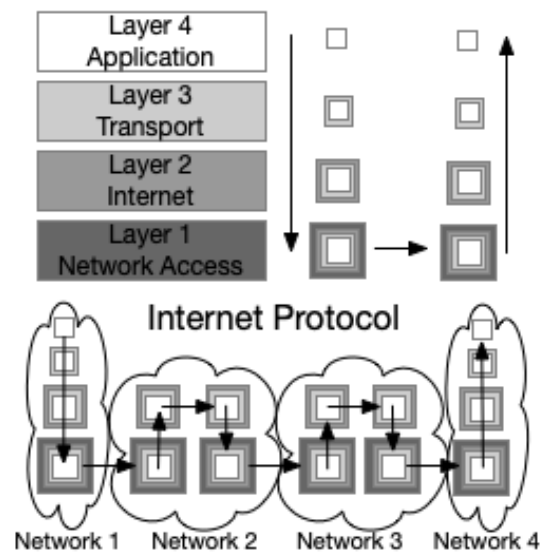

**Figure 2. IP Encapsulation**

When an application is ready to send anything over a network or the Internet, it need only pass that data to the next layer down. Big pieces are broken into smaller pieces as necessary and each one of those pieces is then *encapsulated* – assigned some unique information that will allow it to be reassembled by the corresponding layer at the other end. Note that encapsulation does not include encryption which is a separate process.

Now that the Application layer has encapsulated the smaller pieces, they can be handed off to the lower layers. Then the lower layers take care of their respective network duties, further encapsulating each packet at each layer. Until they are sent over the network and routed to their final destination.

The process is repeated in reverse at the other end. Each layer peels off the encapsulation and passes the data up to the next level until all the data has arrived at the Application Layer of the receiving host. Here, the Application layer reassembles all the smaller pieces back in to a complete image, movie, email or anything else.

This process means that an application programmer does not need to worry about the mechanics of the network layer, and the network engineers don't care what language the applications programmer uses.

There is a guiding principal that each layer should operate independently of the others. There are, however, situations that require one layer to utilize information from another layer. While there is an active effort to minimize this, there are certain security features and other functionality that requires the layers to interact.

Any change of this magnitude requires years of research and consensus from the ISOC's various advisory boards, specifically the Internet Engineering Task Force (IETF) via the Internet Advisory Board (IAB) and the Internet Research Task Force (IRTF), as well as input from many outside sources. One example of an upcoming change of this magnitude is QUIC – a possible new HTML standard (in the Application Layer) developed by engineers at Google, that is based on UDP instead of TCP (in the Transport Layer). All the discussion and development of any code or standards related to QUIC are available to the public to read/comment and/or contribute to in a public Github Repository[10].

## 5. Usage-Based Economic Models

### 5.1. Traditional Analog Usage-Based Billing

To understand basic Internet networking economics, we can start with a comparison to analog phone lines.

Advantage: Because each line was separate, a user was guaranteed a certain level of service once a connection was made.

Disadvantage: If there were too many users and not enough lines, there was no connection – "All Circuits Are Busy."

Disadvantage: The line and its bandwidth are unused when no-one is talking on that line.

Economic Impact: Fixed cost makes upfront budgeting linear, but very susceptible to overuse or underuse leading to either poor user experience or wasted resources.

### 5.2. Digital "Best-Effort" Connections

A digital Best Effort network connection means that the network does not guarantee the delivery of a given packet. It is up to the sender and receiver to confirm that all information was transferred correctly. The Internet is a best effort delivery system, and this has advantages and disadvantages.

Advantage: Shared circuits mean more flexibility in allocating loads.

Advantage: The same networking hardware can be used to carry Voice, Video, Multimedia, and Data.

Disadvantage: If there are too many users and not enough bandwidth, the connection may slow down.

Disadvantage: If the connection slows too much, certain applications may fail. For example: A voice conversation may no longer be intelligible, or video quality may suffer or even stop.

Disadvantage: Both end-points will require computational power for error checking.

### 5.3. Measuring Internet Usage

Because the Internet is 100% digital, all information can be measured in 1s and 0s. Binary.

Usage is measured by how many 1s and 0s can pass through a given point in a certain amount of time.

Each 1 or 0 is a bit. **A bit is the smallest possible unit of measure.**

There are two basic measures of usage: speed and total data.

**Speed** refers to bits per second. The faster a connection, also referred to as higher bandwidth, the more 1s and 0s that can pass through that point in a given second. This measurement takes into account all factors that affect speed at a given point but does not necessarily give a full picture of end-to-end speed. Measured in bits per second (bps, Kbps, Mbps, Gbps).

**Data** refers to the total amount of data to be transmitted regardless of time. Usually measured in Bytes (Byte, KB, MB, GB, TB). 1 Byte = 8 bits.

### 5.4. Basic Internet Billing Structures

**5.4.1 Bandwidth Only**

A customer buys a certain amount of bandwidth. Can be thought of as a pipe of fixed size. The customer is paying for the diameter of the pipe. Home users buy a smaller pipe. Corporate users buy a bigger pipe.

### 5.4.2. Data Only

A customer buys the ability to download a certain amount of data at whatever speeds the network wants to allocate. The pipe diameter can change, and the customer cannot control it. Not a very common solution.

### 5.4.3. Combination of Bandwidth and Data.

A customer agrees to pay a certain amount for guaranteed speed up to a certain amount of data, so the diameter of the pipe is fixed at a larger size, up to a certain amount of data, at which point the pipe may get smaller based on the network's needs or the customer pays a premium for a larger pipe.

### 5.4.4. Dark Fiber

Additional flexibility is gained from Dark Fiber (not Dark Web). Larger customers may use Dark Fiber – fiber that is allocated as shared use. The line is said to remain "dark" until a given customer leases it for a short period of time. The same resources are allocated to a small number of companies to share on a first-come / first-served basis as needed. Dark fiber may be allocated in advance (more expensive) or on the fly (less expensive), however dynamic switching on the fly may result in delay, so you would not want to use dynamic for something like the Superbowl.

### 5.5. Encapsulation, Usage and Billing

None of the aforementioned billing structures needs to address content because of encapsulation.

Any application data is encapsulated before going on the network.

The network engineers and administrators do not need to know what application or content is travelling over the network because it is all encapsulated. And application programmers and engineers don't need to worry about hardware.

Usage can be measured at any point very easily down to the smallest unit – 1 bit.

Currently the Internet Service Provider (ISP), who owns the lines, satellites, and bandwidth carrying the data, bills their customer based on usage – not content.

The ISP sets a price for their bandwidth and it is up to the ISP's customer to determine their individual needs or the needs of their customers. The customer may be a single individual household, a large internet streaming service, or another ISP. The customer decides what is necessary to meet their needs and purchases sufficient bandwidth and data from the ISP.

The Hardware and Networking layers of the Internet by definition, do not care what content that data is carrying.

Further, it is up to the customer to make sure that they have sufficient infrastructure, computational power, and bandwidth to serve their own customer's needs. The ISP is only responsible for their side of the network connection.

### 5.6. Stimulus

As noted in the introduction, it is beyond the scope of this paper to discuss the effects of any stimulus to the equation, however it should be pointed out that various countries including the U.S., most of Europe, Japan, Australia and many others, have employed different methods of supply-side stimulus to encourage the spread of broadband availability and speed improvement – as well as differing methods to affect demand-side consumer price[11].

## 6. Net Neutrality

**Net Neutrality enters the equation when the network providers want to charge based on content.** The debate in the U.S. has moved to the courts and is grounded in a few basic concepts. The basics of the debate are not usually contested, it's the results of any changes and the legal foundations for either preventing or allowing those changes that is cause for debate.

### 6.1. Content Based Subscription Packages

The National ISPs want to be allowed to charge certain Content Providers more than others at one end, *and* charge consumers at the other end for access to that same content. In both cases this might be in addition to any usage fees.

Those opposed to net neutrality regulation argue that ISPs need to be able to charge both consumers and providers based on content because certain content providers use more network resources than others, putting un-acceptable demands on the ISP to provide network resources [12]. They claim that regulation will stifle competition amongst ISPs, slowing the spread of broadband and increasing costs to consumers.

Those in favor of net neutrality legislation argue that the network resources should be allocated and paid for based solely on usage rather than content. Further, they argue that the networks need to remain content neutral in order to ensure true competition, innovation and investment along the "edge" of the internet. Proponents argue that this edge activity is what spurs network growth and user need, not the other way around.

### 6.2. "Fast lanes"

The National ISPs want to be able to throttle the speed or charge extra based on content or content provider to provide "fast lanes."

Opponents of net neutrality legislation argue that the content does not always originate on their network, so they are forced to absorb increased network traffic and allocate resources without seeing any direct benefit.

Proponents of net neutrality argue that the ISPs bill their customers based on a usage plan that measures both speed and total data. The consumer is paying for a certain size pipe. The ISP does not refund money when the pipe is un-used, and the consumer cannot exceed the size of the pipe, so why should the ISP charge the consumer more when they maximize their purchased allocation?

### 6.3. ISP Content Providers

Some of the National ISPs are also Content Providers.

Opponents of net neutrality argue this has no bearing on the previous two changes. They argue that this will not lead to ISPs increasing prices to their competitors, but instead will offer consumers discounted prices for in-house entertainment.

Proponents of net neutrality argue that most of the country is only served by one or two high-speed solutions per location, so allowing an ISP to throttle or surcharge their consumers based on content or creator is monopolistic. Most consumers simply do not have any option to change to. They argue this would lead to less competition and innovation, and higher prices to the consumer for less bandwidth.

### 6.4. International ISPs

The National ISPs want regulation to prevent similar billing structures from the International ISPs down to the National level. The National ISPs argue that National ISPs provide the necessary link to needed customers while International ISPs are basically just networking between National ISPs.

The National ISPs argue they provide the crucial link to the consumers of both networking and application services and in many cases they are the sole provider, so the International ISPs need them to reach the consumers.

The International ISPs argue they need to absorb network traffic from all National ISPs and that some National ISPs use far more resources than others. Further, they argue that very little of the traffic they carry originates on their network – affording them little opportunity to charge at the source.

Proponents of network neutrality in the US would argue that the National ISPs represent themselves as competition to themselves when discussing U.S. Net Neutrality, however they present themselves as the sole provider in many cases when negotiating with the International ISPs.

## 7. Legal History

The basis of all legal challenges – either for or against net neutrality legislation lies in determining which agency will regulate the Internet in the U.S., and then what rules that agency will apply.

### 7.1. 2005 Order

In 2005 the FCC released the first agency guidelines regarding the "open Internet" – the four rules [13]:

"To encourage broadband deployment and preserve and promote the open and interconnected nature of the public Internet, ...
• consumers are entitled to access the lawful Internet content of their choice.
• consumers are entitled to run applications and use services of their choice, subject to the needs of law enforcement.
• consumers are entitled to connect their choice of legal devices that do not harm the network.
• consumers are entitled to competition among network providers, application and service providers, and content providers."

### 7.2. National Cable & Telecommunications Association v. Brand X Internet Services (2005)

These rules were challenged by industry groups representing the ISPs as being too broad. And greatly complicating the issue down the road, in Section II, Paragraph 4, the FCC specifically states that the ISPs are not Title II but are in fact covered under Title I ancillary.

On June 27, 2005 the United States Supreme Court (applying Chevron deference) upheld a determination by the FCC that cable Internet providers were an "information service," and not a "telecommunications service" as classified under the Telecommunications Act of 1996. Supreme Court Justice Antonin Scalia wrote the dissenting opinion, and we will see that his words would later seem prescient when quoted in the majority opinion ten years later in 2016[15].

### 7.3. 2010 Open Internet Order

In a 2010 Executive Order, President Obama directed the FCC to formalize regulations regarding Net Neutrality. The 2010 Order applied much of the same oversight of "common carriers" to the ISPs. The 2010 Order laid out three basic principles: transparency, no blocking, and no unreasonable discrimination.

### 7.4. Verizon Communications Inc. v. FCC (2014)

The 2010 Order was tested in the Supreme Court when Verizon sued the FCC. The Supreme Court ruled that two of the three main principles of the 2010 Order did not fall under the FCC's jurisdiction of Title I ancillary. The Court upheld the Transparency requirements but found that the other two requirements required the FCC to treat ISPs as "common carriers." In its findings, the Court suggested some possible fixes. The Court also recognized the "Virtuous Cycle" of edge providers and recognized that ISPs had the "means and the incentive" to disrupt this cycle to the detriment of consumers.

### 7.5. 2015 REPORT AND ORDER ON REMAND, DECLARATORY RULING, AND ORDER

After a public comments period, the FCC sought to re-define the regulations to define ISPs as "common carriers" and further codify the rules regarding Net Neutrality. In February of 2015 the FCC published the *REPORT AND ORDER ON REMAND, DECLARATORY RULING, AND ORDER*.

The 2015 Order classifies ISPs under Title II of the Communications Acts of 1934,1996[14] and it gives strict guidelines regarding Net Neutrality. The 2015 Order states no blocking, no throttling, and no paid prioritization. The FCC sited the 2014 Verizon findings in the order and pointed to the Court's reference to the "Virtuous Cycle" as authority to change the ISP classification to "common carrier."

### 7.6. United States Telecom Association v. FCC and Independent Telephone and Telecommunications Alliance (2016).

The large telecom companies sued the FCC, however this time the smaller telecoms joined on the side of the FCC. On June 14 of 2016, the U.S. Court of Appeals for the District of Columbia refused to hear the case, thereby upholding the ruling in USTA v FCC that the FCC had operated under its authority to re-classify the ISPs as "common carriers."

Circuit Judge Srinivasan joined by Circuit Judge Tatel wrote the majority opinion. Amongst the various considerations was the following [15]:

> Justice Scalia's dissenting opinion [*in the 2005 decision*], joined by Justices Souter and Ginsburg, went even further. According to Justice Scalia, the statute permitted only one conclusion: cable broadband ISPs "are subject to Title II regulation as common carriers, like their chief competitors [e.g., DSL] who provide Internet access through other technologies." Id. at 1006 (Scalia, J., dissenting). The agency, in Justice Scalia's view, had no discretion to conclude otherwise. And he expressly accepted that his reading of the Act would result in "common carrier regulation of all ISPs," a result he considered "not a worry." Id. At 1011.

The large telecoms appealed to the U.S. Supreme Court and on November 5[th], 2018 the U.S. Supreme Court refused to hear the case, thereby upholding the previous findings.

### 7.7. 2017 - 2018 FCC Changes Direction

In 2017 President Donald Trump appointed Ajit Pai as Chairman of the FCC. Chairman Pai has changed the FCC's direction and decided the FCC should not regulate ISPs with regard to Net Neutrality, thereby rendering the Court's ruling temporarily moot as it applies to the current direction of the FCC. The FCC did stipulate that ISPs must be transparent in any throttling or associated fees.

The Supreme Court's 2018 finding does however mean that should the FCC change direction again, that the stricter rules would have legal precedence. It may also be cited as precedence in any upcoming legal challenges to the current interpretation.

## 8. Researcher Conclusions

The research shows that there is sufficient evidence to support the idea of the "Virtuous Cycle," that it is the activity by innovators, investors, and content providers along the edge of the Internet that drives network growth and innovation, and that this cycle should be protected.

Further, this researcher finds that an economic system based strictly on usage of encapsulated packets is capable of responding to all the needs of the infrastructure market, while ensuring competition along the edge. The research also shows that the reverse is not true - the ISPs have an incentive and the opportunity to abuse their position with regard to competition and this would negatively impact the edge.

The large ISPs have made no secret of their goals. In Verizon Oral Arguments Transcript at line 31 "I'm authorized to state by my client [Verizon] today that, but for these rules, we would be exploring those commercial arrangements…"

Therefore, it is this researcher's recommendation that the principles of encapsulation, capitalism and democracy would best be served by regulation of ISPs to ensure that the basic tenets of Net Neutrality (as defined at the beginning of this paper and codified in the 2015 FCC Order) are followed.  These basic regulations would not restrict any technical decision making by the ISPs, however they would prevent the ISPs from using their often-unique position to restrict access based on content, removing the temptation to start down a slippery slope.

## 9. Areas for Further Exploration

- Net Neutrality at the State and Local Level.
- Legislation affecting local municipalities providing public internet.
- Effects of Net Neutrality on 5G deployment and vice versa.
- The mechanism for resolving inter-ISP disputes.
- The effects of stimulus on supply-side or demand-side economics.

## References

[1] FCC Adopts a Policy Statement Regarding Network Neutrality, Tech Law Journal, 2005, No Author Cited, http://www.techlawjournal.com/topstories/2005/20050805.asp
[2] FCC Passes Compromise Net Neutrality Rules. December 21, 2010. Wired, Sam Gustin. https://www.wired.com/2010/12/fcc-order/
[3] Net Neutrality Explained: What It Means (and Why It Matters). November 23, 2017.  Fortune, Andrew Nusca. http://fortune.com/2017/11/23/net-neutrality-explained-what-it-means-and-why-it-matters/
[4] Net Neutrality in Canada & The Telus Incident Of 2005, December 7, 2017, Sarah Bauer, https://www.navigatormm.com/blog/net-neutrality-in-canada-the-telus-incident-of-2005/
[5] Why is the Internet so Important? Reference.com https://www.reference.com/technology/internet-important-db9d1faab5337241
[6] 47 U.S.C. § 230(b)(2).
[7] 47 U.S.C. § 230(b)(1).
[8] GN Docket No. 17-199 - In the Matter of: Inquiry Concerning Deployment of Advanced Telecommunications Capability to All Americans in a Reasonable and Timely Fashion.  FCC 2018 BROADBAND DEPLOYMENT Report. Adopted: February 2, 2018, Released:  February 2, 2018. By the Commission:  Chairman Pai and Commissioners O'Rielly and Carr issuing separate statements; Commissioners Clyburn and Rosenworcel dissenting and issuing separate statements. Paragraph 15.
[9] RFC 3233 Defining the IETF and 3935 A Mission Statement for the IETF.
[10] https://github.com/quicwg
[11] FCC Global Broadband Deployment Report 2016
[12] Verizon Reply at 63; Letter from Robert C. Barber, AT&T to Marlene H. Dortch, Secretary, FCC, WC Docket No. 10-90. Attach. at 15-19 (filed July 30, 2014) (AT&T July 30, 2014 Ex Parte Letter).
[13] FCC POLICY STATEMENT, 2005,
[14] FCC 2015 REPORT AND ORDER ON REMAND, DECLARATORY RULING, AND ORDER. Paragraph 29.
[15] United States Court of Appeals FOR THE DISTRICT OF COLUMBIA CIRCUIT FILED: MAY 1, 2017. No. 15-1063. UNITED STATES TELECOM ASSOCIATION, PETITIONER v. FEDERAL COMMUNICATIONS COMMISSION AND UNITED STATES OF AMERICA, RESPONDENTS INDEPENDENT TELEPHONE & TELECOMMUNICATIONS ALLIANCE, ET AL., INTERVENORS.

# Appendix A

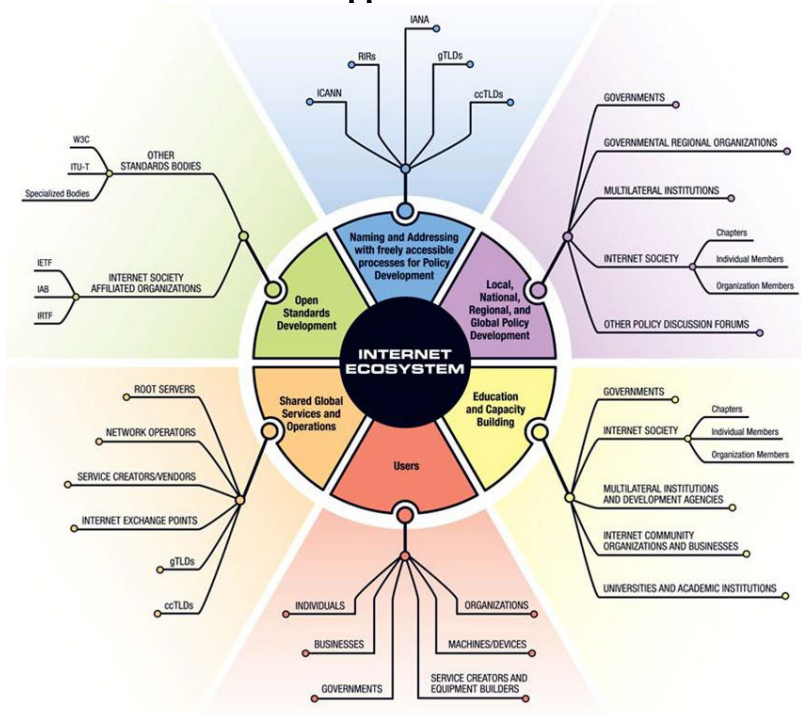

"Internet Ecosystem: naming and addressing, shared global services and operations, and open standards development."
– ISOC March 7, 2014